\documentclass[useAMS,usenatbib]{mn2e}
\usepackage{txfonts}
\usepackage{graphicx,times}
\usepackage{natbib}
\usepackage{amssymb}
\usepackage{longtable}
\usepackage{subfigure}
\title[Mass and Age of Red Giant Branch Stars Observed with LAMOST and \emph{Kepler}]{Mass and Age of Red Giant Branch Stars Observed with LAMOST and \emph{Kepler}}
\author[Yaqian Wu]{Yaqian Wu$^{1}$\thanks{E-mail:
wuyaqian@mail.bnu.edu.cn; bisl@bnu.edu.cn}, Maosheng Xiang$^{2,7}$\thanks{E-mail: msxiang@nao.cas.cn}, Shaolan Bi$^{1}$, Xiaowei Liu$^{3}$, Jie Yu$^{4,5}$, Marc Hon$^{6}$
 \newauthor Sanjib Sharma$^{4}$, Tanda Li$^{4,5,2}$, Yang Huang$^{3,7}$, Kang Liu$^{1}$, Xianfei Zhang$^{1}$, Yaguang Li$^{1}$
 \newauthor Zhishuai Ge$^{1}$, Zhijia Tian$^{3,7}$, Jinghua Zhang$^{1}$, Jianwei Zhang$^{1}$\\
$^{1}$Department of Astronomy, Beijing Normal University,
             Beijing 100875, P.\ R.\ China;\\
$^{2}$National Astronomical Observatories, Chinese Academy of Sciences,
             Beijing 100012, P.\ R.\ China;\\
$^{3}$Department of Astronomy, Peking University,
             Beijing 100871, P.\ R.\ China;\\
$^{4}$School of Physics, University of Sydney,
             Sydney 2000, NSW, Australia;\\
$^{5}$Stellar Astrophysics Centre, Department of Physics and Astronomy,
      Aarhus University, Ny Munkegade 120, DK-8000 Aarhus C, Denmark;\\
$^{6}$School of Physics, The University of New South Wales,
             Sydney 2052, NSW, Australia;\\
$^{7}$LAMOST Fellow
             }
\begin{document}

%\date{Accepted 2017 December 15. Received 2017 December 14; in original form 1988 October 11}

\pagerange{\pageref{firstpage}--\pageref{lastpage}} \pubyear{2017}

\maketitle

\label{firstpage}

\begin{abstract}
Obtaining accurate and precise masses and ages for large numbers of giant stars is of great importance for unraveling the assemblage history of the Galaxy. In this paper, we estimate masses and ages of 6940 red giant branch (RGB) stars with asteroseismic parameters deduced from \emph{Kepler} photometry and stellar atmospheric parameters derived from LAMOST spectra. The typical uncertainties of mass is a few per cent, and that of age is $\sim$\,20 per cent. 
The sample stars reveal two separate sequences in the age -- [$\alpha$/Fe] relation -- a high--$\alpha$ sequence
with stars older than $\sim$\,8\,Gyr and a low--$\alpha$ sequence composed of stars with ages ranging from younger
than 1\,Gyr to older than 11\,Gyr. We further investigate the feasibility of deducing ages and masses directly from LAMOST spectra with a machine learning method based on kernel based principal component analysis, taking a sub-sample of these RGB stars as a training data
set. We demonstrate that ages thus derived achieve an accuracy of $\sim$\,24 per cent. We also explored the feasibility of estimating ages and masses based on the spectroscopically measured carbon and nitrogen abundances. 
 The results are quite satisfactory and significantly improved compared to the previous studies.
\end{abstract}

\begin{keywords}
stars: fundamental parameters -- stars: evolution -- stars: asteroseismology.
\end{keywords}

\section{Introduction}

Accurate age estimates for large numbers of stars are important for a full understanding of the stellar populations and the assemblage history of the Galaxy. Age is one of the key parameters that determine the evolutionary state of a star. However, unlike other parameters, such as mass and chemical composition, direct age estimate for a star from observation or fundamental physical law is extremely difficult if not impossible \citep[e.g.][]{Soderblom10}. Often, one has to rely on stellar evolutionary models for stellar age estimation -- usually achieved by comparing the stellar parameters deduced from the observation with the model predictions.

A practical way of age estimation for individual field stars is to compare their position on the Hertzsprung-Russell diagram (HRD) with the theoretical stellar isochrones, utilizing stellar atmospheric parameters yielded by spectroscopy and/or photometry. The method has been successfully applied to obtain reasonable age estimates for a few thousand F/G type stars with high-resolution spectroscopy \citep{Edvardsson1993, Takeda2007, Haywood13, Bergemann14} or ubvy$\beta$ photometry, compled with precise trigonometric parallax data %combining accurate trigonometric parallax data
 \citep{Nordstrom04}. Recently, Wu et al. (2017) used this method to obtain accurate ages for about 200 main-sequence turn-off stars and used the results to check the accuracy of age estimation for tens of thousand main-sequence turn-off and subgiant stars targeted by the LAMOST spectroscopic surveys \citep{Xiang15a, Xiang17b}. %and applied to the LAMOST spectroscopic surveys to deliver robust age estimates for millions of main sequence turn-off and sub-giant stars \citep{Xiang15a, Xiang17b, Wu17}.
However, the method is difficult to apply to giant stars, as isochrones of giants of different ages are severely crowded together on the HR diagram, particularly in temperature. On the other hand, giant stars are advantaged probes of the Galactic structure as they can be detected to large distances given their high luminosities. Modern large-scale surveys, such as the LAMOST Galactic spectroscopic surveys \citep{Zhao12, Deng12} and the Apache Point Observatory Galactic Evolution Experiment \citep[APOGEE;][]{Majewski10}, have collected high quality spectra that yield robust stellar atmospheric parameters and chemical abundances for millions of giant stars. It has become an emergent task to develop an effective way that delivers robust age estimates for large numbers of giant stars detected by the existing and upcoming large spectroscopic surveys.

Asteroseismologic studies have demonstrated that solar-like oscillations excited by convective turbulence in the stellar envelope are always solar-like stars of 0.8--3.0\,$M_{\odot}$ \citep{Aerts10, Soderblom10, Chaplin13}. Both the grid modelling that fits the individual observed frequencies \citep[e.g.][]{Chaplin14} and a scaling relation based on global asteroseismic parameters, $\Delta\nu$, the frequency separation of two modes of the same spherical degree and consecutive radial order, and $\nu_{\rm max}$, frequency with the maximal oscillation power \citep{Kjeldsen1995}, have been used to infer mass and radius from the oscillation data. Once the precise mass of a giant star has been determined, robust age, to an accuracy of about 20 -- 30 per cent \citep[e.g.][]{Soderblom10, Gai11, Chaplin14}, then can be estimated by comparing the inferred mass with the theoretical stellar isochrones of given metallicity, since the main sequence lifetime of a star is tightly correlated with its mass \citep[e.g.][]{Martig16, Ness2016}. Precise photometry from the \emph{CoRoT} \citep{b1}, \emph{Kepler} \citep{b2} and K2 \citep{Howell14} space missions has allowed determinations of seismic parameters for thousands of red giant stars \cite[e.g.][]{b3,b4,b5,b6,b7,b9,b10,stello17,Mathur16,Yu16}. Many of those stars have also reliable atmospheric parameters yielded by spectroscopy, either from the literature \citep[e.g.][]{Chaplin14,pinsonneault14,Huber13}, or targeted by the APOGEE \citep{Majewski10} or the LAMOST-\emph{Kepler} surveys \citep{b14}, so the ages of these stars can in principle be well-determined and adopted as benchmarks for age estimation for large numbers of giant stars detected in large-scale surveys.

Martig et al. (2016) estimated mass and age for 1475 red giants in the APOKASC sample that have atmospheric parameters and elemental abundances including [C/Fe] and [N/Fe] deduced from APOGEE spectra, as well as seismic parameters derived from \emph{Kepler} photometry (Pinsonneault et al. 2014).
Based on the fact that values of carbon and nitrogen abundance ratio [C/N] of red giant stars is tightly correlated with mass as a consequence of the CNO cycle and the first-dredge up process, they further constructed a relation of stellar mass and age as a function of C and N abundances. Using the APOKASC sample as a training set, \citet{Ness2016} estimated masses and ages for 70,000 APOGEE red giants (including red clump (RC) stars) with $Cannon$, a spectroscopic stellar parameters determination pipeline \citep{Ness2015}. Recently, Ho et al. (2017) applied the method of \citet{Martig16} to LAMOST DR2 and obtained age estimates for 230,000 red giant stars. Nevertheless, we note that due to the relatively large (5 per cent) uncertainties in the asteroseismic parameters of the APOKASC sample stars, typical uncertainties of stellar masses inferred from the asteroseismic measurements are larger than 0.2\,$M_{\odot}$. As a consequence, the resultant age  estimates for the APOKASC sample stars, as well as the afore-mentioned age estimates for the APOGEE and LAMOST giant stars using the APOKASC sample as a training set, have typical uncertainties larger than 40 per cent. Better mass and age estimates of smaller errors are therefore clearly desired. % and it seems that further detailed error clarifications for the above age estimates are desired.

     Recently, Yu et al. (2017) re-derived asteroseismic parameters, including $\Delta\nu$ and $\nu_{\rm max}$, utilizing the four-year \emph{Kepler} data, with typical uncertainties smaller than 3 per cent for red giant stars.
     With these accurate parameters, stellar masses can be inferred to a precision of $\sim$\,7\,per cent.
About 14,000 stars spectra in the sample of Yu et al. were targeted by LAMOST by June, 2016. Precise stellar parameters and elemental abundances, (e.g. [$\alpha$/Fe], [C/H], [N/H]), of those stars were derived from the LAMOST spectra with the LSP3 \citep{Xiang15b, Xiang17a}. Meanwhile, systematic errors of the standard scaling relation were better assessed \citep{White11,Sanjib16,Guggenberger16,Viani17}.%Sharma et al. (2016) have investigated the systematic error of the standard scaling relation and modified the relation by multiplying a correction factor that depends on the evolutionary state, heavy elemental abundance $Z$, effective temperature $T_{\rm eff}$, as well as $\Delta\nu$ and $\nu_{\rm max}$.
With these advantages, it seems that the time is ripe for better mass and age estimates for the large number of giants available from LAMOST surveys. % It is therefore an appropriate time to deliver better ages for LAMOST giant stars.

In this work, we determine masses and ages for the above LAMOST-\emph{Kepler} common red giant branch (RGB) stars. The results are shown to have a median error of 7 per cent in mass and 25 per cent in age estimates. With this LAMOST-\emph{Kepler} sample, we investigated possible correlations between stellar age and chemical composition. We further selected a sub-sample of stars with smaller uncertainties in mass and age estimates as a training set to estimate masses and ages directly from the LAMOST spectra with a machine learning method based on kernel-based principal component analysis (KPCA). Feasibility of the method is discussed, and the resultant age and mass estimates are compared with those inferred from an empirical relation based on the C and N abundances. We found that with the current approach, one can derive stellar ages from the LAMOST spectra with a high precision of 23 per cent.

The paper is organized as follows. Section\,2 introduces the LAMOST-\emph{Kepler} sample. Section\,3 describes the mass and age estimation.
Correlations between the age and elemental abundances are presented in Section\,4. Section\,5 presents ages and masses estimated directly from the LAMOST spectra with the KPCA method. Section\,6 discusses age and mass estimation based on the abundance ratio [C/N], followed by conclusions is Section\,7.

\section{The LAMOST-\emph{Kepler} sample}

\subsection{Atmospheric and asteroseismic parameters}

By June, 2016, the LAMOST-\emph{Kepler} project \citep{b14} collected more than 180,000 low-resolution ($R\sim1800$) optical spectra ($\lambda$\,3800 -- 9000\,${\AA}$) in the \emph{Kepler} field utilizing the LAMOST spectroscopic survey telescope \citep{Cui12}. Robust stellar parameters, including radial velocity $V_{\rm r}$, effective temperature $T_{\rm eff}$, surface gravity log\ $g$, and metallicity [Fe/${\rm H}$], were delivered from the spectra with the LAMOST Stellar Parameter pipeline (LASP; Wu et al. 2011, Luo et al. 2015). Independently, in addition to the above four parameters, interstellar reddening $E_{B-V}$, absolute magnitudes ${\rm M}_V$ and ${\rm M}_{K_{\rm s}}$, $\alpha$-element to metal (and iron) abundance ratio [$\alpha$/M] and [$\alpha$/Fe], as well as carbon and nitrogen abundance [C/H] and [N/H], have also been derived with the LAMOST Stellar Parameter Pipeline at
Peking University \citep[LSP3;][]{Xiang15b, Xiang17b}, utilizing spectra processed with a specific flux calibration pipeline aiming to get better treatment with interstellar extinction of the flux standard stars (Xiang et al. 2015c). Given a spectral signal-to-noise ratio (SNR; 4650\,${\AA}$) higher than 50, stellar parameters yielded by LSP3 have typical precision of about 100\,K for $T_{\rm eff}$, 0.1\,dex for log\ $g$, 0.3 -- 0.4\,mag for $M_{V}$ and $M_{Ks}$, 0.1\,dex for [Fe/${\rm H}$], [C/H] and [N/H], and better than 0.05\,dex for [$\alpha$/M] and [$\alpha$/Fe] \citep{Xiang17c}.

 In this work, we used the global asteroseismic parameters, i.e., $\nu_{\rm max}$ and $\Delta\nu$
     systematically measured by Yu et al. (2017). They used the full-length of four-year \emph{Kepler} time series and a modified version of the SYD pipeline \citep{huber09} to precisely extract
     the seismic parameters. Their results are in good agreement with literatures (e.g. Huber et al. 2011, Hekker et al. 2011, Stello et al. 2013, Huber et al. 2014, Mathur et al. 2016, Yu et al 2017.), displaying a median fractional residual of 0.2 per cent in $\nu_{\rm max}$ with a scatter of 3.5 per cent, and a median fractional residual of 0.01 per cent in $\Delta\nu$ with a scatter of 4.2 per cent.
 %Utilizing the four-year full-length \emph{Kepler} data, Yu et al. (in preparation) have extracted global seismic parameters, including $\Delta\nu$ and $\nu_{\rm max}$ of over 16,000 oscillating red giants.
  Meanwhile, typical uncertainties of the derived seismic parameters are smaller than 3 per cent for red giant stars with log\ $g$ $\gtrsim$ 2.0\,dex. A cross-identification of LAMOST-\emph{Kepler} stars yields 13,504 LAMOST spectra with SNRs higher than 20 for 8654 unique stars.%stars that have LAMOST spectral SNRs higher than 20.

 \subsection{Selection of the RGB stars}
 Considering that RC stars suffer from significant mass loss poorly constrained with the current data and the impact of mass loss on age estimation remains to be investigated in detail, we focus on RGB stars only in this work.
 To pick out RGB stars, we first select stars with $\nu_{\rm max}$ $>$ 120\,$\mu$Hz and $\nu_{\rm max}$ $<$ 12\,$\mu$ Hz, as they are genuine RGB stars. In the regime of 12 $\mu$Hz $<$ $\nu_{\rm max}$ $<$ 120 $\mu$Hz, RGB and RC stars are mixed together, we thus use the results of Hon et al. (2017), who classified the evolutionary state with a machine learning method utilizing the seismic parameters from Yu et al., as well as other results in the literature  that differentiate RGB and RC stars \citep{bedding11,b9,Mosser14,Vrard16,Elsworth17}. Finally, we end up with 6940 RGB stars that have LAMOST spectra.

   The top panel of Fig.\,1 shows the distribution of those RGB sample stars in the HR diagram. The effective temperatures of the sample stars cover the range of 3500 $-$ 5500\,K and the surface gravities vary from 1.5 to 3.3\,dex.
    The bottom panel of Fig.\,1 plots their distributions in the [Fe/${\rm H}$] - [$\alpha$/Fe] plane. The Figure shows two prominent sequences of stars in the plane of chemical composition, which is corresponding to the thin and thick disk sequence, respectively. Above the line are thick disk stars of high [$\alpha$/Fe] and low [Fe/${\rm H}$] values, whereas those below are thin disk stars of low [$\alpha$/Fe] and high [Fe/${\rm H}$] values. The distribution resembles those from the previous high-resolution spectroscopic studies \citep[e.g.][]{Haywood13,Hayden15} or predicted by Galactic chemical modelling \citep[e.g.][]{Schonrich09}. %The high-$\alpha$ sequences, along with a lower-[$\alpha$/Fe] tail in the higher metallicity end, was suggested to belong the thick disk, while the low-$\alpha$ sequence was suggested to belong the thin disk \citep[e.g.][]{Haywood13}.
    Note that there also exists a small number of extremely metal-poor stars with [Fe/${\rm H}$] $<$ -1.0\,dex, probably belong to the halo population.

   \begin{figure}
\centering
\includegraphics[width=90mm, angle=0]{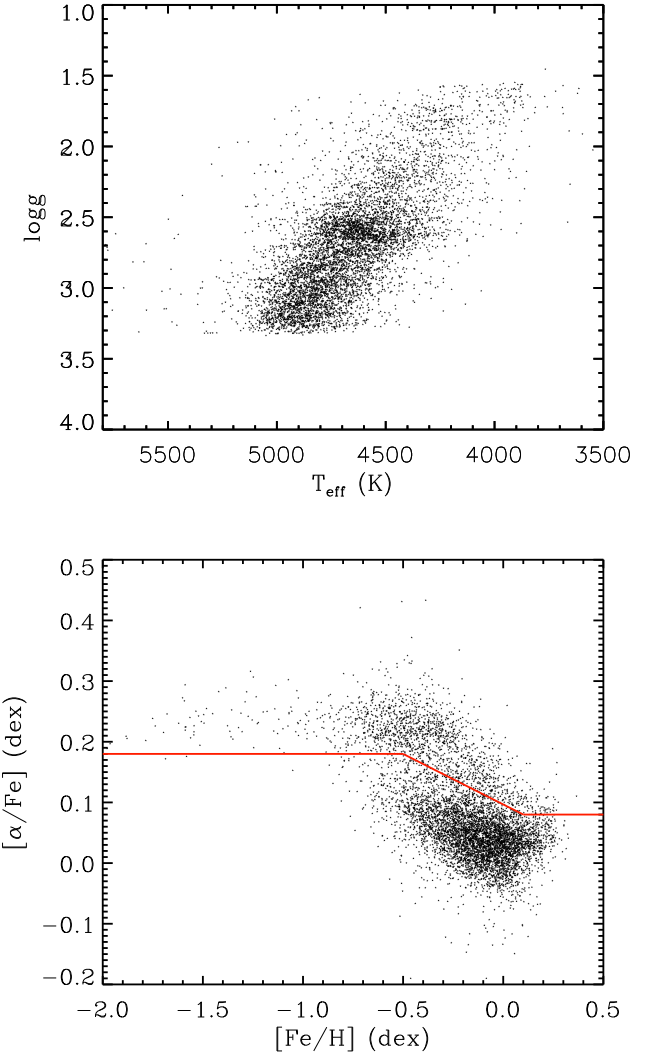}
\caption{Distribution of the RGB sample stars in the $T_{\rm eff}$ -- log\,$g$ (upper) and the [Fe/${\rm H}$] -- [$\alpha$/Fe] (lower) planes. The solid line delineates the demarcation of the thin and thick disk sequences of Haywood et al. (2013). }
\label{Fig:1}
\end{figure}

  \section{Age Determination}
  \subsection{Determining masses with seismic scaling relations}
  Oscillations of solar-like stars are usually described by two global seismic parameters, the frequency of maximum power, $\nu_{\rm max}$, and the mean large frequency separation, $\Delta\nu$. $\nu_{\rm max}$ scales with the acoustic cutoff frequency \citep{Brown91} that depends mainly on surface gravity and temperature \citep{Kjeldsen1995, Belkacem11},
  \begin{equation}\label{1}
    \nu_{\rm max} \propto gT_{\rm eff}^{-1/2} \propto MR^{-2}T_{\rm eff}^{-1/2}.
  \end{equation}
  $\Delta\nu$ is directly related to the travel time of the sound from the centre to the surface of a star, and is therefore sensitive to the mean stellar density \citep{Tassoul1980, Ulrich1986, Kjeldsen1995},
  \begin{equation}\label{0}
    \Delta\nu \propto \rho^{1/2} \propto M^{1/2}R^{-3/2}.
  \end{equation}
   Combining of Equations\,(1) and (2) yields the standard seismic scaling relation that links the stellar mass and radius with seismic parameters and effective temperature,
  \begin{equation}\label{2}
\frac{M}{M\odot}=(\frac{\Delta\nu}{\Delta\nu_{\odot}})^{-4}(\frac{\nu_{\rm max}}{\nu_{\rm max,\odot}})^{3}(\frac{T_{\rm eff}}{T_{\rm eff,\odot}})^{1.5},
  \end{equation}
  \begin{equation}\label{3}
\frac{R}{R\odot}=(\frac{\Delta\nu}{\Delta\nu_{\odot}})^{-2}(\frac{\nu_{\rm max}}{\nu_{\rm max,\odot}})(\frac{T_{\rm eff}}{T_{\rm eff,\odot}})^{0.5}.
  \end{equation}
  We adopt solar values $T_{\rm eff,\odot}$ = 5777\,${\rm K}$, $\nu_{\rm max,\odot}$ = 3090 $\mu$Hz and  $\Delta\nu_{\odot}$ = 135.1 $\mu$Hz \citep{huber11}. It is suggested that uncertainty of the standard scaling relation is about 3 -- 5 per cent for solar-type main-sequence stars and 10 -- 15 per cent for RGB stars (Huber et al. 2011). To reduce the systematic errors, there are several works that attempt to modify the scaling relations \citep{White11,Sanjib16,Guggenberger16,Viani17}. We adopted the modification strategy of Sharma et al. (2016), because it considers in detail the effect on mass, [Fe/${\rm H}$] and $T_{\rm eff}$ (interpolation over a grid of models), and moreover, the code to do this is publicly available. %We adopt the modification strategy of Sharma et al. (2016), because it considers detailedly the mass, [Fe/H], $T_{\rm eff}$, $\Delta\nu$, $\nu_{\rm max}$ and evolutionary state, and also the code is publicly available.
  The standard relations then become:
  \begin{equation}\label{3}
    \frac{M}{M_{\odot}}=(\frac{\nu_{\rm max}}{\nu_{\rm max,\odot}})^{3}(\frac{\Delta\nu}{f_{\Delta\nu}\Delta\nu_{\odot}})^{-4}(\frac{T_{\rm eff}}{T_{\rm eff,\odot}})^{1.5},
  \end{equation}

  \begin{equation}\label{4}
    \frac{R}{R_{\odot}}=(\frac{\nu_{\rm max}}{\nu_{\rm max,\odot}})(\frac{\Delta\nu}{f_{\Delta\nu}\Delta\nu_{\odot}})^{-2}
    (\frac{T_{\rm eff}}{T_{\rm eff,\odot}})^{0.5}.
  \end{equation}
  Here $f_{\Delta\nu}$ is a modification factor.
%The correction factor, $f_{\Delta\nu}$, depends on the evolutionary state, effective temperature, metallicity, $\Delta\nu$ and $\nu_{\rm max}$.
We derived masses for LAMOST-\emph{Kepler} RGB stars with the modified scaling relations, utilizing temperatures from the LSP3 KPCA method, $\Delta\nu$ and $\nu_{\rm max}$ from Yu et al., and $f_{\Delta\nu}$ generated with the Asfgrid code \citep{Sanjib16}. Errors of masses are estimated via propagating errors of $\Delta\nu$, $\nu_{\rm max}$ and $T_{\rm eff}$.%For the uncertainty of the mass, we use the error propagation from the uncertainties on $\Delta\nu$, $\nu_{\rm max}$ and $T_{\rm eff}$ to reach out.

The left panel of Fig.\,2 plots distributions of the derived mass estimates and their percentage errors for the RGB sample stars. Most stars have mass between 0.8 and 1.8\,$M_{\sun}$, and the percentage errors peak at 7 per cent. There are some stars with relatively large ($>40$ per cent) errors, most of them belong to upper RGB that seismic parameters with relatively large uncertainties. %given the short time span of data.
Also over-plotted in the Figure is the histogram of mass errors of the APOKASC sample stars adopted by Martig et al. (2016) and Ness et al. (2016) for age estimation. It shows that the current sample has on average significantly smaller mass errors than those of the APOKASC, which has a median error larger than 0.14\,$M_{\odot}$. However, note that similar to Martig et al. (2016), we ignored mass errors induced by uncertainties of the scaling relations itself. There are some hints that the scaling relations may have considerable uncertainties \citep{Gaulme16}.%all did not consider the influence of uncertainty of scaling relation, it may bring some uncertainty to the results of mass estimated by scaling relation \citep{Gaulme16}.

\begin{figure*}
\centering
\includegraphics[width=180mm, angle=0]{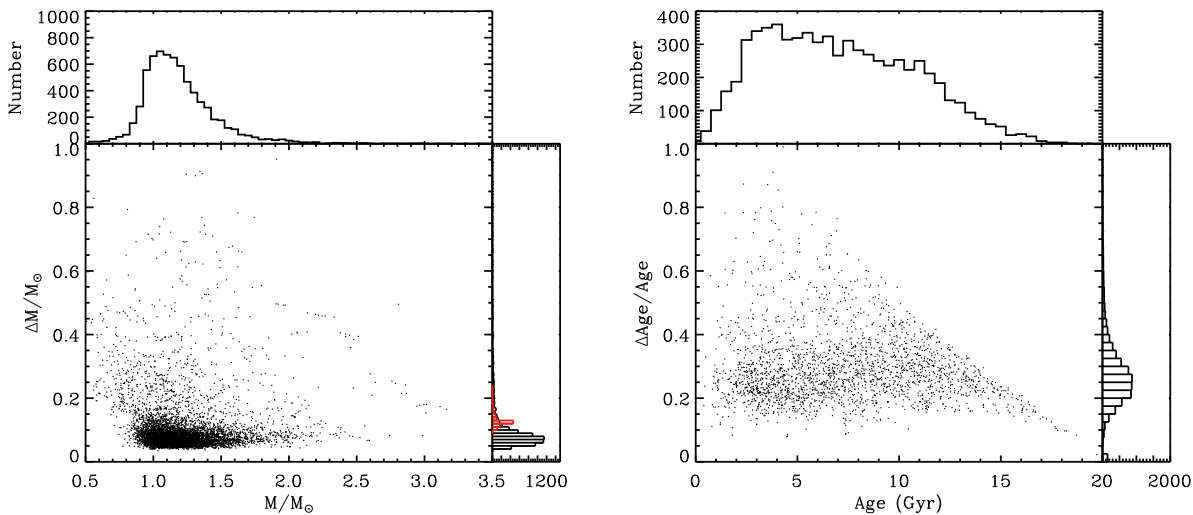}
\caption{Distributions of mass (left) and age (right) estimates, as well as their errors, for the RGB sample stars. The red histogram in the left panel gives the mass error distribution of the APOKASC sample adopted by Martig et al. (2016). }
\label{Fig:2}
\end{figure*}

  \subsection{Estimating ages with isochrones}
  By combining masses inferred from the asteroseismic parameters with stellar atmospheric parameters, ages of RGB stars can be estimated using stellar isochrones. To estimate the stellar age, we adopted a Bayesian method similar to Xiang et al. (2017). The input stellar parameters are masses and log\,$g$ inferred from seismic parameters, and $T_{\rm eff}$, [Fe/${\rm H}$] and [$\alpha$/Fe] derived with the LSP3 KPCA method from LAMOST spectra. Note that we have corrected the LSP3 effective temperatures to the scale as given by the color-metallicity-temperature relation for giant stars of Huang et al. (2015). The Yonsei-Yale (Y$^2$) isochrones (Demarque et al. 2004) were adopted because the database cover a wide range of age, [Fe/${\rm H}$] and [$\alpha$/Fe]. The Dartmouth Stellar Evolution Database (DESP; Dotter et al. 2008) and the PAdova and TRieste Stellar Evolution Code (PARSEC; Bressan et al. 2012) were also used in order to examine the uncertainty of age estimation induced by the stellar models. Comparing the results of different isochrones, we found that the age estimated with the DSEP isochrones are comparable to those estimated with the Y$^2$ isochrones, with a mean difference of 6 per cent and a dispersion of 8 per cent. Ages estimated with the PARSEC isochrones are also consistent well with those estimated with the Y$^2$ isochrones. Note that all the three isochrones did not consider effects from stellar rotation, metal diffusion and magnetic fields, which may have some impact on the age determination.%have not significant distinction compared with Y$^2$ for most of the stars. However, at present, we did not consider the influence of rotation and magnetic field of star, it may cause some effect on determining ages of star.

  As an example of the age estimation, Fig.\,3 plots the posterior probability distributions as a function of age for three stars of typical ages. The existence of prominent and well-defined peaks of the distributions suggest that the resultant ages are well constrained. The robustness of our age estimation benefits to a large extent from the precise mass estimation from seismology. Distributions of estimated ages and errors for the whole RGB sample are plotted in the right panel of Fig.\,2. The sample covers the whole range of possible ages of stars, from close to zero on the young end, up to the age of the universe ($\sim$13.8\,Gyr; Planck collaboration 2016). There are still a small number of stars with unphysical ages, i.e. older than 13.8\,Gyr. This is most likely the consequence of parameter errors in either asteroseismic or atmospheric. Typical relative age errors are 15 per cent--35 per cent, with a median value of 25 per cent. A small fraction ($\sim$ 5 per cent) of stars exhibit large age errors ($>50$ per cent). Again this is mainly due to large uncertainties in mass estimates and/or in stellar atmospheric parameters.

   \begin{figure*}
\centering
\includegraphics[width=\textwidth, angle=0]{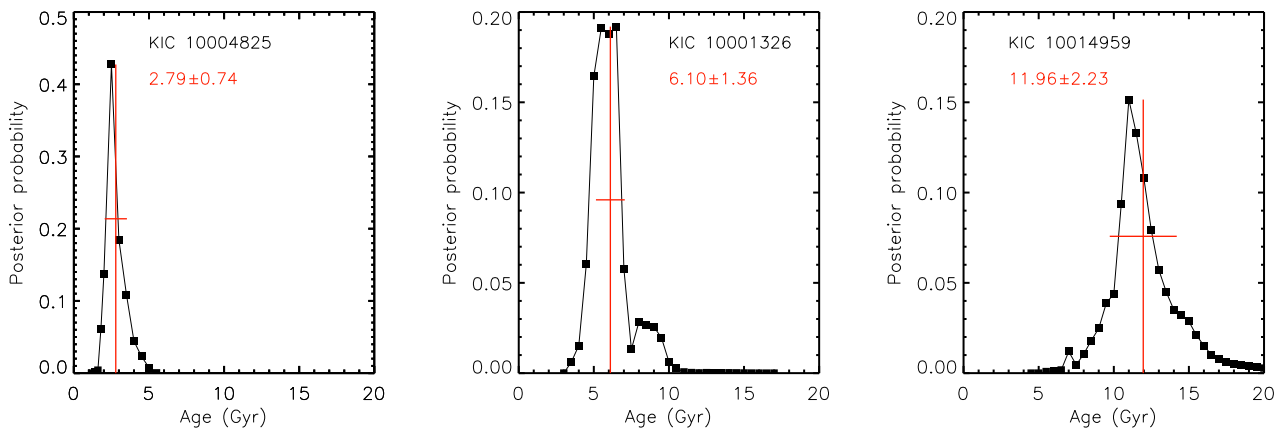}
\caption{Posterior probability distributions for age estimation for three example stars of different ages. The vertical and horizontal lines in red denote the mean ages and errors, respectively, whose values are marked in the Figure.  }
\label{Fig:4}
\end{figure*}

 Sample entries of all parameters deduced for the RGB sample are listed in Table.\,1. In addition to the deduced parameters, $T_{\rm eff}$, log\ $g$, [Fe/${\rm H}$], [$\alpha$/Fe], [C/N], mass and age, additional information about the stars, including the KIC ID, are given. The sample entries of all parameters are public available on-line.%the Right Assension RA and Declination Dec, are given. %Full table for the whole sample is available onine\footnote{http://...}%The derived masses and ages of the RGB sample stars, along with their atmospheric parameters, are listed in Table.\,1.
 % Among the 6940 RGB stars, the stellar parameters are successfully derived, the sample table are listed in Table.\,1, including thd information about KIC, RA, Dec, $T_{\rm eff}$, log\ $g$, [Fe/${\rm H}$], [$\alpha$/Fe], [C/N], mass and age.

\begin{table*}
\caption{Sample entries of deduced parameters of the 6940 RGB sample stars.}
%\caption{Best fit coefficients for mass for red giant stars as a quadratic function of $T_{\rm eff}$, log\ $g$, [Fe/${\rm H}$], [$\alpha$/Fe], [C/N] and [C/Fe].}\label{tab:coef_mass2}
\begin{tabular}{cccccccc}
\hline
  Star & ${T_{\rm eff}}$ & $\log g$ & $[\rm{Fe/H}]$  & [$\alpha$/Fe] & $[\rm {C/N}]$ & Mass & Age \\

  KIC & (K) & (dex) & (dex)  & (dex) & (dex) & $M_{\odot}$ & (Gyr) \\																																							
\hline
     7866696  & $4277\pm147$ & $2.36\pm0.01$ & $-0.21\pm0.1$ & $0.06\pm0.03$ & $-0.2\pm0.1$ & $1.12\pm0.12$ & $8.2\pm3.1$ \\
     7728741  & $4480\pm160$ & $3.03\pm0.01$ & $0.05\pm0.13$ & $0.04\pm0.05$ & $-0.19\pm0.1$ & $1.00\pm0.08$ & $13.1\pm3.6$ \\
     7728945  & $4900\pm125$ & $2.87\pm0.01$ & $-0.23\pm0.14$ & $0.01\pm0.05$ & $-0.3\pm0.1$ & $1.28\pm0.09$ & $4.7\pm1.3$ \\
     10382554  & $4560\pm111$ & $2.72\pm0.01$ & $-0.3\pm0.09$ & $0.22\pm0.03$ & $0.02\pm0.1$ & $0.99\pm0.06$ & $12.5\pm2.9$ \\
     8004863 &  $4813\pm80$ & $2.89\pm0.01$ & $-0.15\pm0.12$ & $0.05\pm0.03$ & $-0.16\pm0.1$ & $1.32\pm0.07$ & $3.3\pm0.4$ \\
     7936033  & $4075\pm152$ & $2.03\pm0.02$ & $-0.14\pm0.14$ & $0.23\pm0.03$ & $-0.09\pm0.1$ & $0.90\pm0.13$ & $5.2\pm0.7$ \\
     7798102  & $4561\pm99$ & $2.55\pm0.01$ & $-0.48\pm0.1$ & $0.24\pm0.03$ & $0.02\pm0.1$ & $1.02\pm0.12$ & $9.2\pm2.0$ \\
    ...\\
    \hline
    \end{tabular}
    %\end{tabular}
    \end{table*}

   \section{Correlations among Mass, Age, metallicity and abundance}

   Utilizing this sample of RGB stars, we explore possible correlations among the mass, age, metallicity and elemental abundances. To ensure high data quality, stars with age errors larger than 40 per cent or mass errors larger than 15 per cent were discarded, leaving 3726 unique stars in the remaining sample.

   Fig.\,4 plots the distributions of the median stellar ages and masses in the [Fe/${\rm H}$] -- [$\alpha$/Fe] and [Fe/${\rm H}$] -- [C/N] planes. From the Figure, it shows clear variations of the median age with [Fe/${\rm H}$] and  [$\alpha$/Fe]. For a given [Fe/${\rm H}$], stars of higher [$\alpha$/Fe] have older ages. There is an old ($>10$\,Gyr) sequence of stars on the high-$\alpha$ side, spanning continuously in [Fe/${\rm H}$] from a value of $\lesssim-1.0$\,dex to a super-solar metallicity. The trajectory of this old sequence of stats in the [Fe/${\rm H}$] -- [$\alpha$/Fe] plane agrees well with the high-$\alpha$, thick disk sequence of stars seen in Fig.\,1. The thin disk sequence in the abundance plane are dominated by young and intermediate-age stars, respectively. The patterns observed here are generally consistent with those of Ho et al. (2017). In particular, Fig.\,5 of Ho et al. also shows that stars in the low-$\alpha$ tail of the thick disk sequence are as old as their high-$\alpha$, metal-poor counterparts, although this is less clear in their results due to the much larger uncertainties of their age estimates than ours. The current results however deviate to some extent from those of Xiang et al. (2017) and Haywood et al. (2013), who find slightly younger ages of stars in the low-$\alpha$ tail of the thick disk sequence. The reasons for this discrepancy are not fully understood. We suspect they are either simply due to the uncertainties in the age estimates, or caused by the possible population effects, as our sample consists mainly  stars of the inner disk, while the samples of Xiang et al. and Haywood et al. are more mixed, mostly located in the solar neighbourhood or the outer disk. Although the sample of Ho et al. (2017) has also a broad spatial distribution, their age estimates are based on the APOKASC sample, which, as in our case, contains mainly stars in the inner disk. %The difference may lead to some possible some systematics amongst these samples.

 In the [Fe/${\rm H}$]--[C/N] plane, stars with higher [C/N] tend to have older ages for a given [Fe/${\rm H}$], likely a consequence of the CNO cycle. The age distribution of stars in the plane are however does not follow clear sequences as in the [Fe/${\rm H}$]--[$\alpha$/Fe] plane, as stars with higher [C/N] values also have much larger age dispersions.
Stellar masses exhibit clear variations with abundances in both the [Fe/${\rm H}$] -- [$\alpha$/Fe] and the [Fe/${\rm H}$] -- [C/N] planes. Stars with lower values of metallicity, higher [$\alpha$/Fe] or [C/N] values have generally lower masses, a consequence of stellar evolution but the target selection effects may also play a role  -- the cuts in  $T_{\rm eff}$ and log\,$g$ used to select RGB sample stars tend to discard more massive stars of older ages.

Fig.\,5 plots the distributions of stars in the age -- [$\alpha$/Fe], age -- [Fe/${\rm H}$] and age -- [C/N] planes. The distribution in the age -- [$\alpha$/Fe] plane exhibits two prominent sequences of different [$\alpha$/Fe] values. The high-$\alpha$ sequence is mainly composed of old ($>8$\,Gyr) stars that have an almost flat age -- [$\alpha$/Fe] relation. The low-$\alpha$ sequence contains stars with a wide range of ages, from younger than 1\,Gyr to older than 12\,Gyr. Stars older than $\sim$6\,Gyr tend to have a flat age -- [$\alpha$/Fe] relation, while for younger stars, the [$\alpha$/Fe] values decrease with decreasing stellar ages. Such a double sequence distribution is largely in agreement with the finding of Xiang et al. (2017), but the current sample seems to contain a higher fraction of old ($> 10$\,Gyr), low-$\alpha$ stars than that of Xiang et al. (2017), and this difference is clearly responsible for the differences seen in the age distributions of the two samples in the [Fe/${\rm H}$]--[$\alpha$/Fe] plane as discussed above. The current sample also include 42 young ($<5$\,Gyr) stars of unexpectedly high [$\alpha$/Fe] values ($>0.15$\,dex). Such young, high-$\alpha$ stars are also found in  previous work \citep[e.g.][]{Martig15,Chiappini15}. Further studies are needed to understand their origins. One possible explanation is that they were formed near the ends of the Galactic bar (Chiappini et al. 2015). It is also suggested that those stars are actually evolved blue stragglers \citep{Jofre16}.

There is no strong correlation between age and [Fe/${\rm H}$].  At any given age, stars exhibit a wide range of [Fe/${\rm H}$] values. Such a flat age -- [Fe/${\rm H}$] relation is consistent with the previous findings for solar neighbourhood stars \citep[e.g.][]{Nordstrom04,Bergemann14}. It is probably a combination of the consequences of the mixing process that mixes stars born at various positions with different [Fe/${\rm H}$] \citep{Roskar08,Schonrich09,Loebman11} and of the complex star formation and chemical enrichment history of the Galactic disk. The result however differ in some aspect from the finding of some of the previous studies e.g. Haywood et al. (2013), Xiang et al. (2017), who find a tight age -- [Fe/${\rm H}$] correlation for stars older than 8\,Gyr. Compared to the latter two studies, the current sample exhibits an unexpectedly large fraction of old ($>12$\,Gyr), metal-rich ($\gtrsim-0.2$\,dex) stars. We suspect that those stars were born in the inner disk. As discussed above, we believe that the differences are due to the population effects. The distribution of stars in the age -- [C/N] plane shows that older stars tend to have larger [C/N] in general, suggesting some correlation between [C/N] ratio and mass and age. However, for a given age, the dispersion of stars in [C/N] is significant.

 \begin{figure*}
 \centering
 \includegraphics[width=0.9\textwidth]{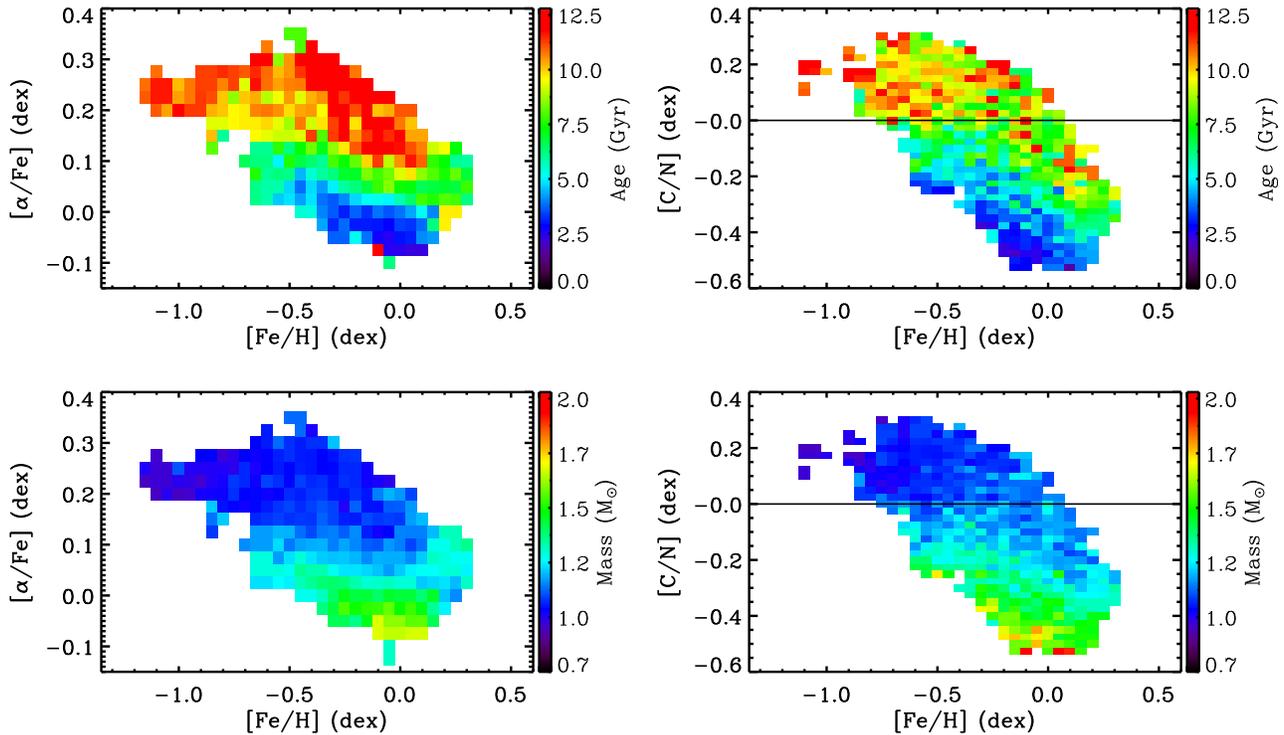}

 \caption{$Upper panels$: Distributions of median stellar ages of [$\alpha$/Fe] -- [Fe/${\rm H}$] (left) and [C/N] --  [Fe/${\rm H}$] (right).
$Lower panels$: Distributions of median stellar masses of [$\alpha$/Fe] -- [Fe/${\rm H}$] (left) and [C/N] --  [Fe/${\rm H}$] (right). The color bar represents the median mass and age of each bin size.
%The thick and thin disk sequences are distinguished by whether they lie above or below the three segments of the black solid line respectively.
A unified bin size of $0.05\times0.025$\,dex is adopted for all the panels.}
 \label{fig:subfig7} %% label for entire figure
\end{figure*}

\begin{figure*}
\centering
\includegraphics[width=\textwidth]{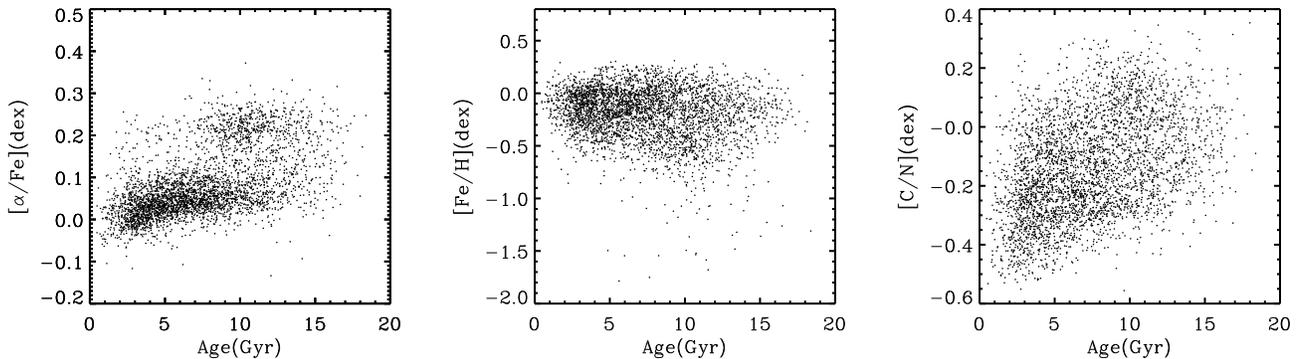}
\caption{Age -- [$\alpha$/Fe] (left), age -- [Fe/${\rm H}$] (middle) and age -- [C/N] relations for the RGB sample stars.}
\label{fig:subfig8}
\end{figure*}

\section{Estimating ages and masses from spectra with the KPCA method}
In this Section, we explore a machine learning method based on kernel based principal component analysis \citep[KPCA;][]{Sch98} to estimate ages and masses directly from the LAMOST spectra. The method has been successfully used to estimate stellar atmospheric parameters \citep{Xiang17a}. We refer to Xiang et al. for a detailed introduction of how to apply the method to LAMOST spectra. Briefly, KPCA is a non-linear method to extract principal components from high-dimensional data sets. It works like PCA but the latter is a linear method working in the data space, whereas KPCA works in a feature space generated by the kernel function. Here we adopted the Gaussian radial basis function. A multiple linear model between the extracted principal components and the target parameters (age, mass) was constructed with a regression method utilizing the training data sets. Parameters of the target spectra were then estimated with the model from their principal components.

\subsection{The training sample}
Sample stars with age errors smaller than 40 per cent or mass errors smaller than 15 per cent, as selected to study the age--metallicity relation in Section\,4, are defined to be the training sample. The sample contains 3726 stars, with a median mass error of 7 per cent and a median age error of 20 per cent. Although in the current work we focus on age and mass estimate for giant stars, we found that keeping a sufficient number of sub-giant and dwarf stars in the training sample is necessary in order to reduce the systematic errors caused by the boundary effects of the method (Xiang et al. 2017a). We have therefore added to the training sample another 314 sub-giant and dwarf stars selected from the LAMOST-\emph{Kepler} common star database that have seismic parameters available from the catalog of Chaplin et al. (2014). In total, the training sample contains 4040 stars with precise seismic mass and age estimates. Fig.\,6 shows the distribution of stars in the training sample in the $T_{\rm eff}$ -- log\ $g$ plane. Effective temperatures of the sample stars range from about 4000 to 6500\,K, and log\,$g$ values range from $\sim$2.0 to 4.5\,dex. There are few stars of $\log~g < 2.0$\,dex, probably because these stars have relatively large uncertainties of seismic parameters.%an artifact resulted from the short length of time series of the $Kepler$ photometry.

\begin{figure}
\centering
%\subfigure{
%\label{fig:subfig9:a}
\includegraphics[width=90mm]{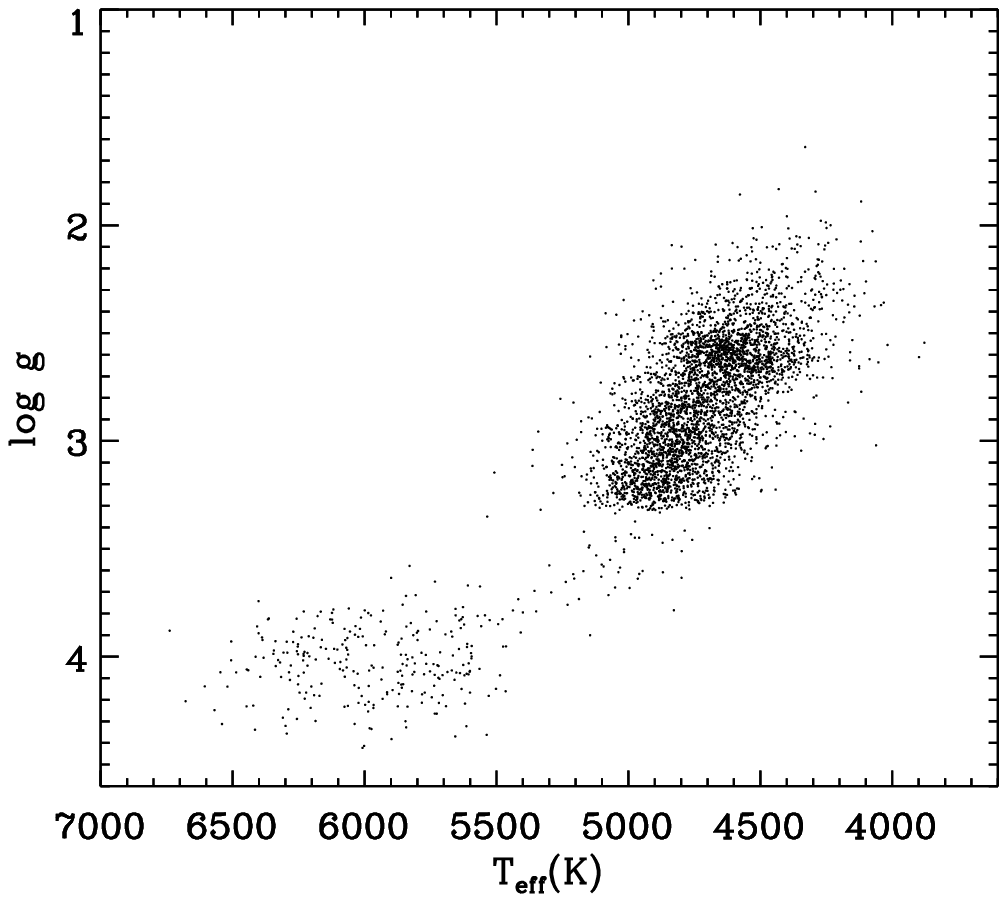}
%\hspace{1in}
%\subfigure{
%\label{fig:subfig9:b}
%\includegraphics[width=90mm]{logg_feh_all.eps}}
\caption{Distribution of the LAMOST¨C-\emph{Kepler} training sample in the $T_{\rm eff}$ -- log\ $g$ planes.}
\label{Fig:subfig9}
\end{figure}
\subsection{KPCA results}
Similar to Xiang et al. (2017a), we adopted the 3900 -- 5500\,${\AA}$\, spectral segment for age and mass determination. To find an optimal value of the number of principal components, $N_{\rm PC}$, we tried various values from 20 to 900, and examined how parameter residuals between the KPCA estimates and the seismic values vary with $N_{\rm PC}$. Generally, a large number of $N_{\rm PC}$ generates smaller residuals between the KPCA estimates and seismic parameters of the training sample. As $N_{\rm PC}$ increases from 20 to 900, the dispersion of the residuals decreases from 26 per cent to 14 per cent for the age, and from 9 per cent to 5 per cent for the mass. However, a large value of $N_{\rm PC}$  has the risk of over-fitting. An excessively large $N_{\rm PC}$ is also improper as the results become very sensitive to the spectral noises/imperfections, leading to poor robustness for spectra of relatively low signal-to-noise ratios (SNRs).

 To determine the optimal $N_{\rm PC}$, we carried out two experiments. One is to divide the sample stars into two groups of equal number, leave-half-out (k-fold-cv), one group is adopted as the training set to estimate ages and masses of the other, test group of stars. Fig.\,7 plots the dispersion of the relative residuals of age for both the training and test groups as a function of $N_{\rm PC}$. It shows that as the $N_{\rm PC}$ increases from 20 to 900, the dispersion of the training sets decreases contiguously from 28 to 12 per cent, whereas the dispersion of the test sets becomes nearly flat when $N_{\rm PC}$ is larger than 100, indicating that the method may suffer from over-fitting at such large number of principal components. This also suggests 100 is an optimal value of $N_{\rm PC}$. Another experiment is the so-called leave-one-out cross-validation (LOOCV) -- the sample stars (of number $N$) are divided into two groups, a training group containing $N-1$ stars and a test group containing one single star. The age and mass of the test star are estimated with the KPCA method taking the training group as the training set. The exercise is repeated $N$ times to estimate the ages and masses of all the sample stars. Fig.\,7 illustrates that the method generates slightly smaller dispersion than that of test sample of first experiment. This is largely caused by the different size of the training sets, as the LOOCV method uses $N-1$ stars while the k-fold-cv method uses $N/2$ stars. %This is a little in conflict with the first experiment. The reason is not fully understood, possibly because the test spectra have some sort of correlations with the training spectra due to common systematic errors.
For safety, we adopt a value of 100 for $N_{\rm PC}$ in this work.

%\begin{figure}
%\centering
%\includegraphics[width=90mm, angle=0]{age-mass1.eps}
%\caption{Comparison of ages and masses yields by the KPCA method and the seismic values for different numbers of principal components. The number of principal components, as well as the standard deviation of the differences, are marked in each panel.}
%\label{Fig:6}
%\end{figure}

\begin{figure}
\centering
\includegraphics[width=90mm, angle=0]{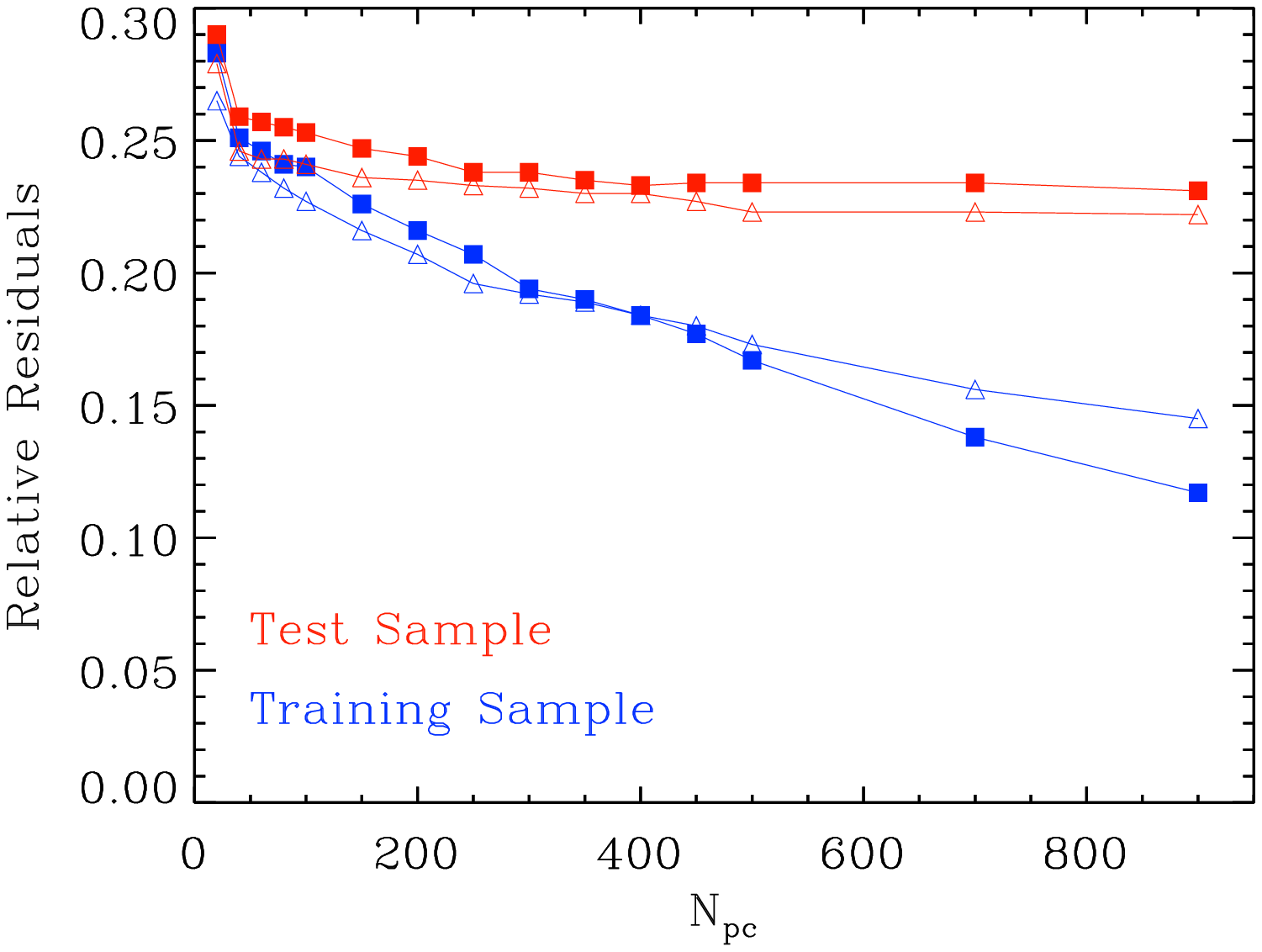}
\caption{The dispersion of the relative residuals of age for both the training and test groups as a function of $N_{\rm PC}$. Red represents test sample, blue represents training sample. The solid square and triangle represent the results of two groups and leave one out respectively.}
\label{Fig:6}
\end{figure}

Fig.\,8 plots the residuals of ages and masses deduced from spectral fitting with the KPCA method as compared to the seismic values as a function of metallicity for the 4040 training sample stars. The Figure shows no significant biases of age and mass estimates for [Fe/${\rm H}$] values down to about $-1.5$\,dex.

To further examine the feasibility of the method, we plot the the median residuals of ages and masses as derived with spectral fitting the seismic values in the $T_{\rm eff}$ -- log\ $g$ plane in Fig.\,9. In general, the Figure exhibits rather homogeneous distribution for both age and mass residuals across the $T_{\rm eff}$ -- $\log\,g$ plane except for some regions near the boundaries. For sub-giant stars of $T_{\rm eff}\sim5200$\,K and $\log\,g\sim3.5$\,dex, the KPCA method may yielded ages systematically higher than the seismic values by $\sim$3\,Gyr, and masses lower than the seismic values by $\sim$0.2\,dex. The reasons are not fully understood. One possibility is a defect in the KPCA method due to small number of sub-giant stars in the training sample.

The method have been applied to the whole spectra set of the LAMOST Galactic surveys, yielding ages and masses for hundreds of thousands of RGB stars. A careful analysis of the huge data set, as well as a further calibration of age estimates with member stars of open clusters, will be presented in a separate work.
%An application of the method to the whole spectra set of the LAMOST Galactic surveys can be delivered ages and masses for hundreds of thousands of RGB stars. A careful discussion and test for open cluster will be on the next work.

\begin{figure}
\centering
%\subfigure[LAMOST - $\emph{Kepler}$ RGB sample]{
% \label{fig:subfig1:a} %% label for first subfigure
 \includegraphics[width=0.6\textwidth]{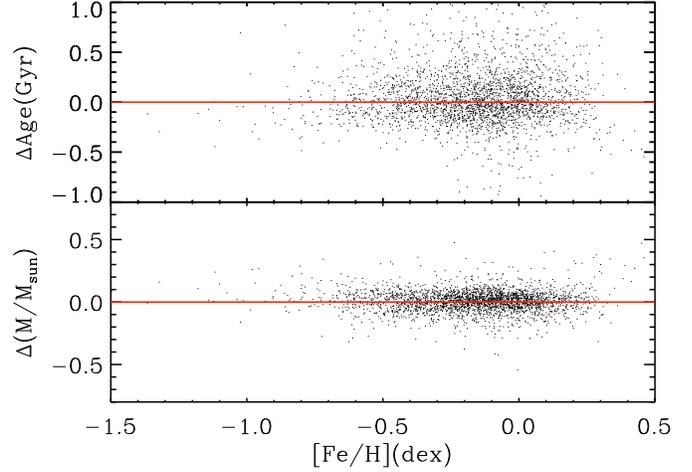}
 %\hspace{1in}
 %\subfigure[LAMOST - $\emph{Kepler}$ stars]{
 %\label{fig:subfig1:b} %% label for second subfigure
 %\includegraphics[width=0.5\textwidth]{teff_age_feh_all.eps}}
 \caption{Variations of the regression residuals of age and mass estimates as a functions of [Fe/${\rm H}$].}
 \label{fig:subfig1} %% label for entire figure
\end{figure}

\begin{figure*}
\centering
 %\subfigure{
 %\label{fig:subfig3:a} %% label for first subfigure
 %\includegraphics[width=0.5\textwidth]{teff_logg_age_lt5.ps}}
 %\hspace{1in}
 %\subfigure{
 %\label{fig:subfig3:b} %% label for second subfigure
 \includegraphics[width=0.9\textwidth]{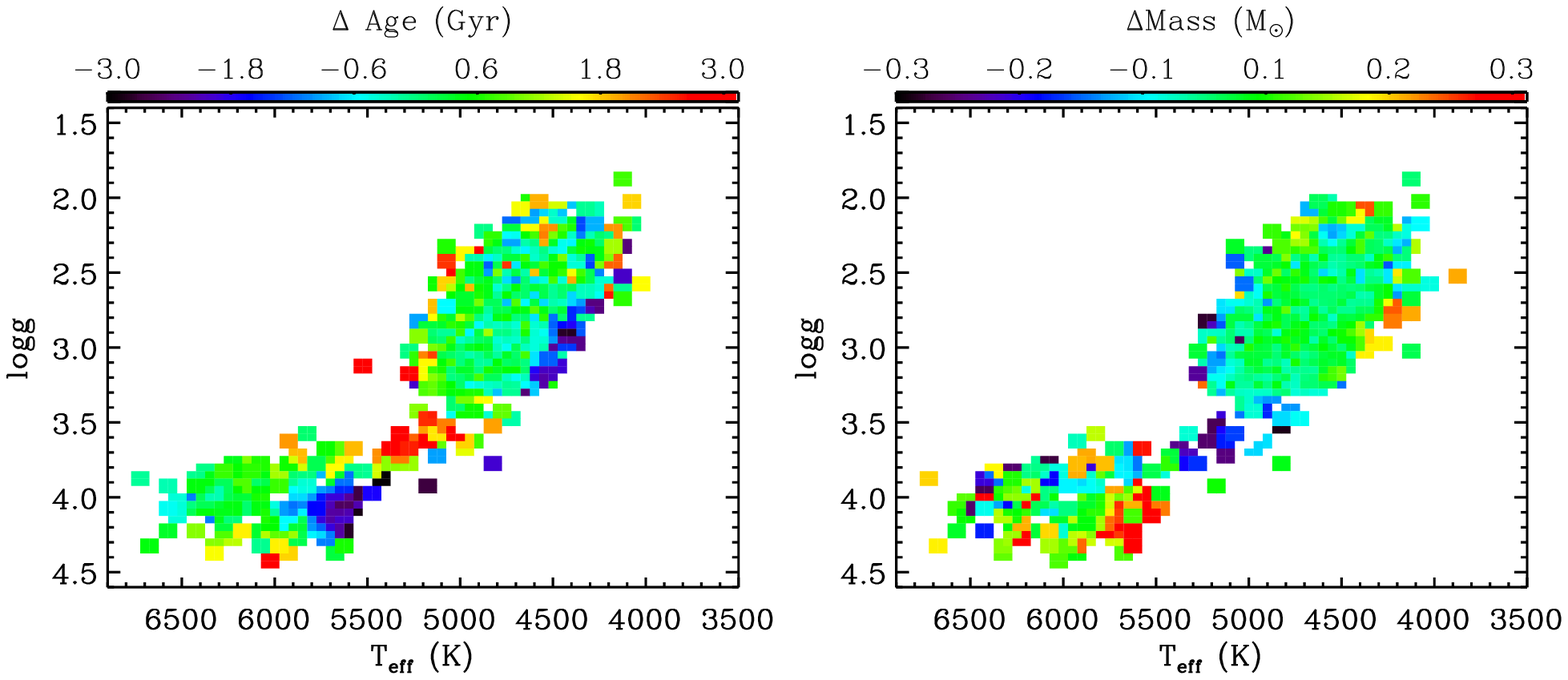}
 \caption{Distributions of residuals of ages (left) and masses (right) estimated with KPCA method as compared to the seismic values in the $T_{\rm eff}$ -- log\,$g$ planes.}
 \label{fig:subfig3} %% label for entire figure
\end{figure*}

\section{Estimating ages and masses based on the carbon and nitrogen abundances}

The observed C and N abundance ratio [C/N] of a RGB star depends on its mass as a consequence of the first dredge-up process, that brings the inner material processed by the CNO cycle out to the stellar surface. This is because [C/N] ratio in the core, as well as the depth of the convective mixing that drives the dredge-up process, depends sensitively on the stellar mass (Masseron \& Gilmore 2015). The [C/N] is thus a good indicator of the mass and age of a RGB star (e.g. Masseron \& Gilmore 2015, Martig et al. 2016).

Following Martig et al. (2016), we estimated ages and masses based on the [C/H] and [N/H] abundances utilizing a polynomial regression method.
We assumed a quadratic function between the mass (age) and the stellar parameters. Two sets of stellar parameters were employed, one includes {$T_{\rm eff}$, log\,$g$, [Fe/${\rm H}$], [$\alpha$/Fe], [C/N] and [C/Fe]}, the other includes {$T_{\rm eff}$, log\,$g$, [Fe/${\rm H}$], [C/N] and [C/Fe]}:
 \begin{equation}\label{11}
   ${\rm Mass (Age)}$ = f(T_{\rm eff}, \log\ $g$, [{\rm Fe}/{\rm H}], [\alpha/{\rm Fe}], [{\rm C}/{\rm N}], [{\rm C}/{\rm Fe}]),
\end{equation}
\begin{equation}\label{12}
  ${\rm Mass (Age)}$ = f(T_{\rm eff},\log\ $g$, [{\rm Fe}/{\rm H}], [{\rm C}/{\rm N}], [{\rm C}/{\rm Fe}]).
\end{equation}

 Least-square fitting algorithm is adopted to determine the coefficients of the quadratic function, utilizing a sample of 3726 stars with age errors smaller than 40 per cent and mass errors smaller than 15 per cent, as defined in Section\,4. Coefficients of the fits are listed in Tables 2 and 3 for the two sets of stellar parameters, respectively.

Fig.\,10 plots a comparison of masses deduced with the fitting formulae Eqs.\,(7) and (8)and the seismic values. The Figure shows good agreement, with a dispersion of the residuals of only 7 per cent and 8 per cent for the two parameter sets with and without [$\alpha$/Fe]. The results are much more precise than a precision of 20 per cent obtained by Martig et al. (2015). Near either the lower or higher mass end, there are the some visible differences. They are likely to be mainly caused by the random errors in seismic mass estimates, although some uncertainties introduced by the poor fitting near the mass boundaries may also contribute some of the differences. A similar comparison for age estimates is shown in Fig.\,11. The residuals have a dispersion of 25 per cent and 26 per cent for the two parameter sets with and without [$\alpha$/Fe]. However, some systematic patterns in the residuals are seen as a function of age. Ages estimated with the carbon and nitrogen abundances are larger than the seismic values by $\sim$\,10 per cent for stars younger than $\lesssim$5\,Gyr, and lower than the seismic values by $\sim$\,10 per cent for stars older than $\gtrsim$10\,Gyr, likely caused by the inadequacy of the regression method. Figs.\,10 and 11 also show that including [$\alpha$/Fe] in the fitting does not significantly improve the precision of mass and age estimation. This is probably because [C/Fe] correlates well with [$\alpha$/Fe].

\begin{table*}
\caption{Best fit coefficients for mass estimation based on the carbon and nitrogen abundances}
%\caption{Best fit coefficients for mass for red giant stars as a quadratic function of $T_{\rm eff}$, log\ $g$, [Fe/${\rm H}$], [$\alpha$/Fe], [C/N] and [C/Fe].}\label{tab:coef_mass2}
\begin{tabular}{c}
 a),Best fit coefficients for mass estimation as a quadratic function of $T_{\rm eff}$, log\ $g$, [Fe/${\rm H}$], [$\alpha$/Fe], [C/N] and [C/Fe] [see Eq.\,(7)]\\

\end{tabular}
\begin{tabular}{cccccccc}
\hline
 & 1 & log\ $T_{\rm eff}$ & log\ $g$ & [Fe/H] & [$\alpha$/Fe] & [C/N] & [C/Fe]\\
\hline
1  & 1138.72   & -625.59   & -2.66   & 9.97   & 37.69   & 37.55 & 4.22   \\
log\ $T_{\rm eff}$  &  & 86.12  & 1.17   & -3.1   & -10.15   & -11.25 & -1.49  \\
log\ $g$ &  &  & -0.33 & 0.54  & -0.14   & 1.07  & 0.43  \\
$\mathrm{[Fe/H]}$& &  &    & -0.24  & -0.65  & -0.5  & -0.34 \\
$\mathrm{[\alpha/Fe]}$ & & &   &  &  -0.15  & 1.52 & 1.03  \\
$\mathrm{[C/N]}$ & &  &  & &  & 0.29 & 0.05   \\
$\mathrm{[C/Fe]}$ & &  &  & &  &  & -0.25 \\
\hline
\end{tabular}
%\end{center}
%\end{table*}

%\begin{table*}
%\begin{center}
%\caption{Best fit coefficients for mass for red giant stars as a quadratic function of $T_{\rm eff}$, log\ $g$, [Fe/${\rm H}$], [C/N] and [C/Fe].}\label{tab:coef_age1}
\begin{tabular}{c}
b),Best fit coefficients for mass estimation as a quadratic function of $T_{\rm eff}$, log\ $g$, [Fe/${\rm H}$], [C/N] and [C/Fe] [see Eq.\,(8)]\\

\end{tabular}
\begin{tabular}{ccccccc}
\hline
& 1 & log\ $T_{\rm eff}$ & log\ $g$ & [Fe/H]& [C/N] & [C/Fe]\\
\hline
1  & 109.11   & -30.84  & -50.28   & 21.22   & 34.49   & 92.02  \\
log\ $T_{\rm eff}$&   & 0.06   & 14.65  & -6.36   & -10.19   & -26.39  \\
log\ $g$&  &    & -0.69   & 0.85   & 0.81   & 1.72   \\
$\mathrm{[Fe/H]}$& & &    & -0.39   & -0.38   & -1.35   \\
$\mathrm{[C/N]}$&  & & &    & 0.50  & -0.01   \\
$\mathrm{[C/Fe]}$ & &  &  & &  & -1.04   \\
\hline
\end{tabular}
%\end{center}
\end{table*}

\begin{figure*}
 \centering
 \includegraphics[width=150mm]{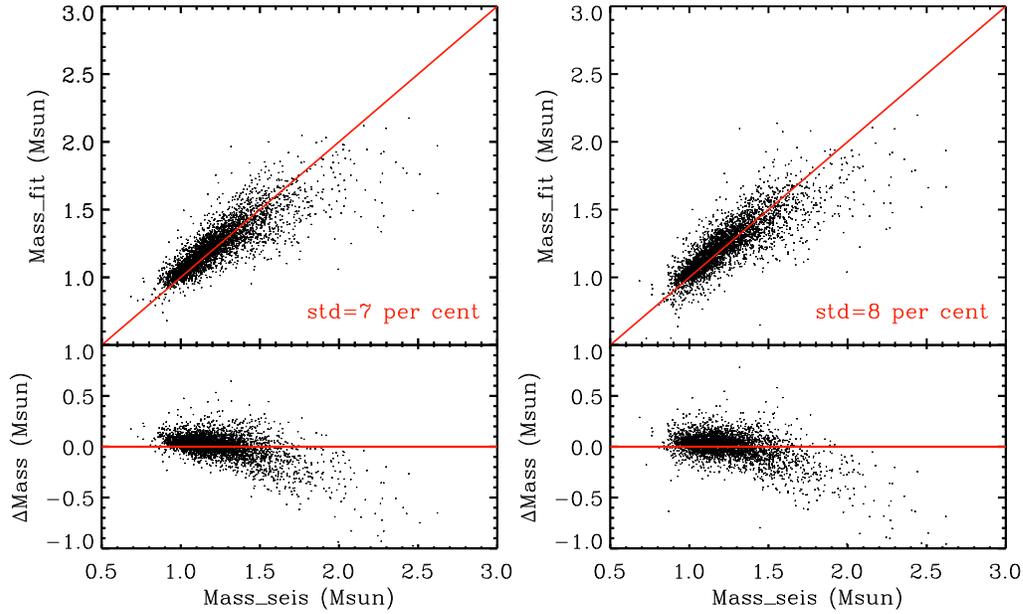}
 \caption{Comparison of masses deduced based on the carbon and nitrogen abundances using fits Eq.\,(7) (left) and Eq.\,(8) (right). %The left figure shows results using {$T_{\rm eff}$, log\,$g$, [Fe/${\rm H}$], [$\alpha$/Fe], [C/N], [C/Fe]} for the fitting method, while the right figure shows results using {$T_{\rm eff}$, log\,$g$, [Fe/${\rm H}$], [C/N], [C/Fe]} for the fitting method.
 The bottom panels show the residuals, with the resistant estimate of the standard deviation marked in red.}
 \label{fig:18} %% label for entire figure
\end{figure*}

\begin{figure*}
 \centering
 \includegraphics[width=150mm]{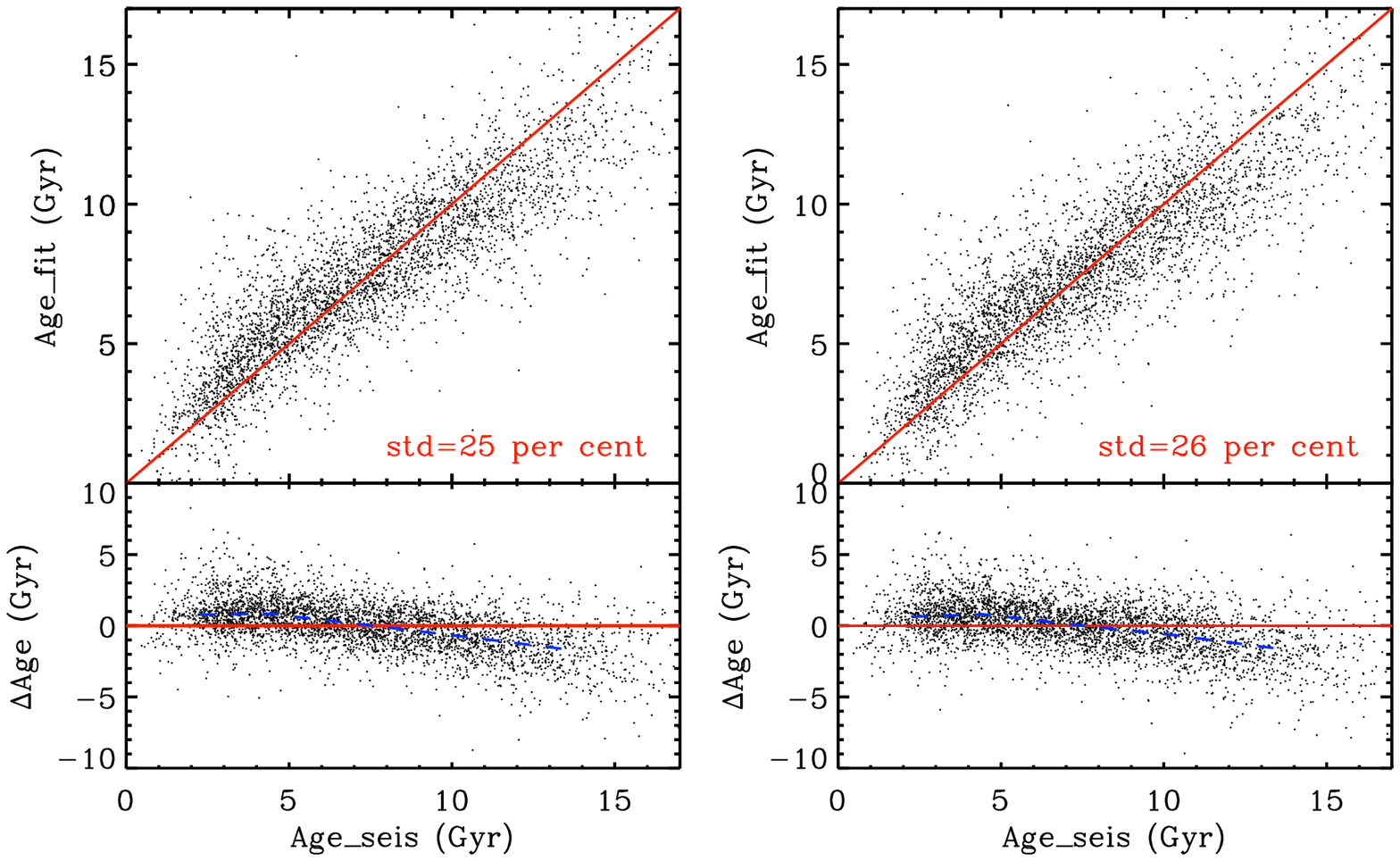}
 \caption{Same as Fig.\,10, but for age estimation.}
 \label{fig:20} %% label for entire figure
\end{figure*}

\section{Conclusion}
We have estimated masses and ages for 6940 RGB stars observed by the LAMOST-\emph{Kepler} project that have accurate asteroseismic parameters. Typical uncertainty is $\sim$7 per cent for the mass estimates, and $\sim$25 per cent for the age estimates. Utilizing a subsample of 3726 RGB stars with small age and mass uncertainties, we have investigated possible correlations among age, [Fe/${\rm H}$], [$\alpha$/Fe] and [C/N]. The results show that the  age -- [$\alpha$/Fe] relation exhibits two separate sequences, a flat, high-$\alpha$ sequence dominated by stars older than $\sim$8\,Gyr, and a low-$\alpha$ sequence composed of stars with a wide range of ages, from younger than 1\,Gyr to older than 13\,Gyr. At any given age, stars exhibit a broad [Fe/${\rm H}$] distribution, leading to a weak age -- [Fe/${\rm H}$] correlation. Particularly, there is a considerable fraction of metal-rich ($\gtrsim-0.2$\,dex) stars of very old ages ($>$10\,Gyr).

Taking the sample stars as a training data set, we show that precise ages and masses can be estimated directly from the LAMOST spectra with a KPCA based machine-learning method. Typical uncertainties of masses and ages thus estimated are comparable to those of the seismic estimates. The method can be applied to the whole database of the LAMOST Galactic Spectroscopic Surveys, yielding robust age and mass estimates for hundreds of thousands of RGB stars.
We have also explored the feasibility of estimating masses and ages from the stellar atmospheric parameters, including the carbon and nitrogen abundances, and found that the method is capable of yielding masses with a uncertainty of better than 10 per cent, as well as age with a uncertainty of better than 23 per cent.

\begin{table*}
%\begin{center}
\caption{Best fit coefficients for age estimation based on the carbon and nitrogen abundances}
\begin{tabular}{c}
a),Best fit coefficients for age estimation as a quadratic function of $T_{\rm eff}$, log\ $g$, [Fe/${\rm H}$], [$\alpha$/Fe], [C/N] and [C/Fe] [seen Eq.\,(7)]\\
\end{tabular}
%\caption{Best fit coefficients for age for red giant stars as a quadratic function of $T_{\rm eff}$, log\ $g$, [Fe/${\rm H}$], [$\alpha$/Fe], [C/N] and [C/Fe].}\label{tab:coef_age}
\begin{tabular}{cccccccc}
\hline
 & 1 & log\ $T_{\rm eff}$ & log\ $g$ & [Fe/H] & [$\alpha$/Fe] & [C/N] & [C/Fe]\\
\hline
1  & -9749.25   & 5842.99   & -109.99   & 603.26   & 607.98   & 1039.59 & 262.19   \\
log\ $T_{\rm eff}$  &  & -789.07  & 27.34   & -167.04   & -181.39   & -282.58 & -73.01  \\
log\ $g$ &  &  & 2.71  & 3.85  & 20.1   & 3.14  & 2.28  \\
$\mathrm{[Fe/H]}$& &  &    & -2.76  & -32.43  & -5.15  & -3.93 \\
$\mathrm{[\alpha/Fe]}$ & & &   &  & 7.93   & -16.39 & -14.46   \\
$\mathrm{[C/N]}$ & &  &  & &  & -2.88 & -1.01   \\
$\mathrm{[C/Fe]}$ & &  &  & &  &  & -4.69 \\
\hline
\end{tabular}
%\end{center}
%\end{table*}
%\begin{table*}
%\begin{center}
\begin{tabular}{c}
 b),Best fit coefficients for age estimation as a quadratic function of $T_{\rm eff}$, log\ $g$, [Fe/${\rm H}$], [C/N] and [C/Fe] [seen Eq.\,(8)]\\
\end{tabular}
%\caption{Best fit coefficients for age for red giant stars as a quadratic function of $T_{\rm eff}$, log\ $g$, [Fe/${\rm H}$], [C/N] and [C/Fe].}\label{tab:coef_age1}
\begin{tabular}{ccccccc}
\hline
& 1 & log\ $T_{\rm eff}$ & log\ $g$ & [Fe/H]& [C/N] & [C/Fe]\\
\hline
1  & -10318   & 6017.42  & -184.05   & 537.77   & 1236.82   & 287.04  \\
log\ $T_{\rm eff}$&   & -873.87   & 50.84  & -151.71   & -340.27   & -82.19  \\
log\ $g$&  &    & 0.83   & 5.84   & 8.45   & 5.58   \\
$\mathrm{[Fe/H]}$& & &    & 0.40   & -9.79   & -1.44   \\
$\mathrm{[C/N]}$&  & & &    & -2.92  & -1.07   \\
$\mathrm{[C/Fe]}$ & &  &  & &  & -5.11   \\
\hline
\end{tabular}
%\end{center}
\end{table*}

\vspace{7mm} \noindent {\bf Acknowledgments}
{This work is supported by the Joint Research Fund in Astronomy (U1631236) under cooperative agreement between the National Natural Science Foundation of China (NSFC) and Chinese Academy of Sciences (CAS), and grants 11273007 and 10933002 from the National Natural Science Foundation of China, and the Fundamental Research Funds for the Central Universities and Youth Scholars Program of Beijing Normal University.
This work is also supported by National Key Basic Research Program of China 2014CB845700
and Joint Funds of the National Natural Science Foundation of China (Grant No. U1531244).
M.-S. Xiang and Y. Huang acknowledge supports from NSFC Grant No. 11703035.
The LAMOST FELLOWSHIP is supported by Special Funding for Advanced Users, budgeted and administrated by Center for Astronomical Mega-Science, Chinese Academy of Sciences (CAMS). T. L. acknowledges funding from an Australian Research Council DP grant DP150104667, the Danish National Research Foundation (Grant DNRF106), grants 11503039 and 11427901 from the National Natural Science Foundation of China.
Guoshoujing Telescope (the Large Sky Area Multi-Object Fiber Spectroscopic Telescope LAMOST)
is a National Major Scientific Project built by the Chinese Academy of Sciences.
Funding for the project has been provided by the National Development and Reform Commission.
LAMOST is operated and managed by the National Astronomical Observatories, Chinese Academy of Sciences.}

{}

\label{lastpage}


\begin{thebibliography}{}
\bibitem[\protect\citeauthoryear{Aerts et al.}{2010}]{Aerts10} Aerts Conny, Christensen-Dalsgaard J{\O}rgen, Kurtz Donald W., 2010, aste.book, A
\bibitem[\protect\citeauthoryear{Baglin}{2006}]{b1} Baglin A., Auvergne M., Barge P., Deleuil M., Catala C., Michel E., Weiss W., in Fridlund M., Baglin A., Lochard J., Conroy L. eds ESA SP1306:The CoRoT Mission Pre-Launch Status - Stellar Seismology and Planet Finding. ESA, Noordwijk, p. 165
\bibitem[\protect\citeauthoryear{Bedding et al.}{2010}]{b6} Bedding T. R. et al., 2010, ApJ, 713, L176
\bibitem[\protect\citeauthoryear{Bedding et al.}{2011}]{bedding11} Bedding T. R., et al. 2011, Nature, 471, 608
\bibitem[\protect\citeauthoryear{Belkacem et al.}{2011}]{Belkacem11} Belkacem K., Goupil M. J., Dupret M. A., Samadi R., Baudin F., Noels A., Mosser B., 2011, A\&A, 530, A142
\bibitem[\protect\citeauthoryear{Bergemann et al.}{2014}]{Bergemann14} Bergemann M. et al., 2014, A\&A, 565, A89
\bibitem[\protect\citeauthoryear{Bressan et al.}{2012}]{Bressan12} Bressan Alessandro, Marigo Paola, Girardi L\'{e}o, et al., 2012, MNRAS, 427, 127B
\bibitem[\protect\citeauthoryear{Brown et al.}{1991}]{Brown91} Brown T. M., Gilliland R. L., Noyes R. W., Ramsey L. W., 1991, ApJ, 368, 599
\bibitem[\protect\citeauthoryear{Borucki et al.}{2010}]{b2} Borucki W. J. et al., 2010, Science, 327, 977
\bibitem[\protect\citeauthoryear{Chaplin \& Miglio}{2013}]{Chaplin13} Chaplin, William J., Miglio, Andrea, 2013, ARA\&A, 51, 353C
\bibitem[\protect\citeauthoryear{Chaplin et al.}{2014}]{Chaplin14} Chaplin W. J., Basu, S., Huber, D., et al., 2014, ApJS, 210, 1
\bibitem[\protect\citeauthoryear{Chiappini et al.}{2015}]{Chiappini15} Chiappini C., Anders F., Rodrigues T. S., et al., 2015, A\&A, 576, 12
\bibitem[\protect\citeauthoryear{Cui et al.}{2012}]{Cui12} Cui X. Q., Zhao Y. H., Chu Y. Q., et al., 2012, RAA, 12, 1197
\bibitem[\protect\citeauthoryear{De Ridder et al.}{2009}]{b3} De Ridder J. et al., 2009, Nature, 459, 398
\bibitem[\protect\citeauthoryear{De Cat et al.}{2015}]{b14} Peter D. C., Fu Jianning, Ren Anbing, et al., 2015, ApJL, 220, 19
\bibitem[\protect\citeauthoryear{Demarque et al.}{2004}]{Demarque04} Demarque P., Woo J.-H., Kim Y.-C., Yi S. K., 2004, ApJS, 155, 667
\bibitem[\protect\citeauthoryear{Deng et al.}{2012}]{Deng12} Deng L. C. et al., 2012, RAA, 12, 735
\bibitem[\protect\citeauthoryear{Dotter et al.}{2008}]{Dotter08} Dotter A., et al. 2008, ApJs, 178, 89
\bibitem[\protect\citeauthoryear{Edvardsson et al.}{1993}]{Edvardsson1993} Edvardsson, B., Andersen, J., Gustafsson, B, et al., 1993, A\&A, 275, 101E
\bibitem[\protect\citeauthoryear{Elsworth et al.}{2017}]{Elsworth17} Elsworth Yvonne, Hekker Saskia, Basu Sarbani, et al., 2017, MNRAS, 466, 3344
\bibitem[\protect\citeauthoryear{Gai et al.}{2011}]{Gai11} Gai N., Basu S., Chaplin W. J., Elsworth Y., 2011, ApJ, 730, 63
\bibitem[\protect\citeauthoryear{Gaulme et al.}{2016}]{Gaulme16} Gaulme P., McKeever J., Jackiewicz J., et al., 2016, ApJ, 832, 121
\bibitem[\protect\citeauthoryear{Goldreich \& Keeley}{1977}]{b12} Goldreich P., Keeley D. A., 1977, ApJ, 212, 243
\bibitem[\protect\citeauthoryear{Guggenberger et al.}{2016}]{Guggenberger16} Guggenberger Elisabeth, Hekker Saskia, Basu Sarbani, et al. 2016, MNRAS, 460, 4277
\bibitem[\protect\citeauthoryear{Hayden et al.}{2015}]{Hayden15} Hayden Michael R., Bovy Jo, Holtzman Jon A., et al., 2015, ApJ, 808, 132
\bibitem[\protect\citeauthoryear{Haywood et al.}{2013}]{Haywood13} Haywood M., Di Matteo P., Lehnert M. D., Katz D., G\'{o}mez A., 2013, A\&A, 560, A109
\bibitem[\protect\citeauthoryear{Hekker et al.}{2009}]{b4} Hekker S. et al., 2009, A\&A, 506, 465
\bibitem[\protect\citeauthoryear{Hekker et al.}{2011}]{b5} Hekker S. et al., 2011, MNRAS, 414, 2594
\bibitem[\protect\citeauthoryear{Hou et al.}{2016}]{Hou16} Hou J. L., Zhong J., Chen L., Yu J. C., Liu C., Deng L. C., 2013, in Wong T., Ott J., eds, Proc. IAU Symp. 292, Molecular Gas, Dust, and Star Formation. Cambridge Univ. Press, Cambridge, p. 105
\bibitem[\protect\citeauthoryear{Ho et al.}{2017}]{Ho17} Ho A. Y. Q., Rix H. W., Ness M. K., Hogg D. W., Liu C., Ting
Y. S., 2017, ApJ, 841, 40H
\bibitem[\protect\citeauthoryear{Howell et al.}{2014}]{Howell14} Howell Steve B., Sobeck, Charlie, Haas, Michael, et al. 2014, PASP, 126, 398H
\bibitem[\protect\citeauthoryear{Huang et al.}{2015}]{huang15} Huang Y., Liu X. W., Yuan H. B., et al. 2015, MNRAS, 454, 2863
\bibitem[\protect\citeauthoryear{Huber et al.}{2009}]{huber09} Huber D., Stello D., Bedding T. R., et al. 2009, CoAst, 160, 74
\bibitem[\protect\citeauthoryear{Huber et al.}{2011}]{huber11} Huber D., Bedding T. R., Stello D., et al., 2011, ApJ, 743, 143
\bibitem[\protect\citeauthoryear{Huber et al.}{2013}]{Huber13} Huber D. et al., 2013, ApJ, 767 127H
\bibitem[\protect\citeauthoryear{Jofr\'{e} et al.}{2016}]{Jofre16} Jofr\'{e} P., Jorissen A., Van Eck S., et al., 2016, A\&A, 595, 60
%\bibitem[\protect\citeauthoryear{J{\o}rgensen \& Lindegren}{2005}]{Jorgensen05} J{\o}rgensen B. R., Lindegren L., 2005, A\&A, 436, 127
%\bibitem[\protect\citeauthoryear{Liu et al.}{2014}]{Liu14} Liu X. W., Yuan H. B., Luo Z. Y., et al., 2014, in IAU Symposium 298 eds. S. Feltzing, G. Zhao, N. A. Walton, \&P. Whitelock, 310
\bibitem[\protect\citeauthoryear{Liu Zhao \& Hou}{2015a}]{Liu15} Liu X.-W., Zhao G., Hou J.-L., 2015a, RAA, 15, 1089
\bibitem[\protect\citeauthoryear{Luo et al.}{2015}]{Luo15} Luo A.-L. et al., 2015, RAA, 15, 1095
\bibitem[\protect\citeauthoryear{Loebman et al.}{2011}]{Loebman11} Loebman S. R., Ro\v{s}kar R., Debattista V. P., Ivezi\'{c}\v{Z}., Quinn T. R., Wadsley J., 2011, ApJ, 737, 8
\bibitem[\protect\citeauthoryear{Kjeldsen \& Bedding}{1995}]{Kjeldsen1995} Kjeldsen H., Bedding T. R., 1995, A\&A, 293, 87
\bibitem[\protect\citeauthoryear{Majewski et al.}{2010}]{Majewski10} Majewski S. R., Wilson J. C., Hearty F., Schiavon R. R., SkrutskieM. F., 2010, in IAU Symposium, Vol. 265, IAU Symposium, Cunha K., Spite M., Barbuy B., eds., pp. 480¨C481
\bibitem[\protect\citeauthoryear{Hon Marc}{2017}]{Marc} Hon Marc, Stello D., Yu Jie., 2017, MNRAS, 469, 4578
\bibitem[\protect\citeauthoryear{Martig et al.}{2015}]{Martig15} Martig M., Rix H. W., Silva Aguirre V., et al., 2015 MNRAS, 451, 2230
\bibitem[\protect\citeauthoryear{Martig et al.}{2016}]{Martig16} Martig M., Fouesneau M., Rix H. W., et al., 2016, MNRAS, 456, 3655
\bibitem[\protect\citeauthoryear{Mathur et al.}{2016}]{Mathur16} Mathur S., Garc\'{\i}a R. A., Huber D., et al., 2016, ApJ, 827, 50M
\bibitem[\protect\citeauthoryear{Miglio et al.}{2013b}]{b11} Miglio A. et al., 2013b, in Montalbn J., Noels A., Van Grootel V., EPJ Web Conf., 43, 40th L\`{\i}ege International Astrophysical Colloquium, Ageing Low Mass Stars: From Red Giants to White Dwarfs. L\`{\i}ege, Belgium, 03004
\bibitem[\protect\citeauthoryear{Mosser et al.}{2010}]{b7} Mosser B. et al., 2010, A\&A, 517, A22
\bibitem[\protect\citeauthoryear{Mosser et al.}{2014}]{Mosser14} Mosser B., Benomar O., Belkacem K., et al., 2014, A\&A, 572, 5M
\bibitem[\protect\citeauthoryear{Ness et al.}{2015}]{Ness2015} Ness M., Hogg David W.,  Rix H.-W., et al., 2015, ApJ, 808, 16N
\bibitem[\protect\citeauthoryear{Ness et al.}{2016}]{Ness2016} Ness M., Hogg David W., Rix H.-W., et al. 2016, ApJ, 823, 114N
\bibitem[\protect\citeauthoryear{Nordstr\"{o}m et al.}{2004}]{Nordstrom04} Nordstr\"{o}m B. et al., 2004, A\&A, 418, 989
\bibitem[\protect\citeauthoryear{Pinsonneault et al.}{2014}]{pinsonneault14} Pinsonneault, Marc H., Elsworth,Yvonne, Epstein, Courtney, et al., 2014, ApJS, 215, 19P
\bibitem[\protect\citeauthoryear{Planck Collaboration et al.}{2016}]{Planck16} Planck Collaboration et al., 2016, A\&A, 594, A13
\bibitem[\protect\citeauthoryear{Ren et al.}{2016}]{Ren16} Ren, J. J., Liu, X. W., Xiang, M. S., et al., 2016, RAA, 16c, 9R
\bibitem[\protect\citeauthoryear{Ro\v{s}kar et al.}{2008}]{Roskar08} Ro\v{s}kar R., Debattista V. P., Quinn T. R., Stinson G. S., Wadsley J., 2008, ApJ, 684, L79
%\bibitem[\protect\citeauthoryear{Samadi \& Goupil}{2001}]{b13} Samadi R., Goupil M.-J., 2001, A\&A, 370, 136
\bibitem[\protect\citeauthoryear{Sharma et al.}{2016}]{Sanjib16}Sharma Sanjib, Stello Dennis, Bland-Hawthorn Joss, Huber Daniel, Bedding Timothy R., 2016, ApJ, 822, 15S
\bibitem[\protect\citeauthoryear{Sharma et al.}{2017}]{Sanjib17} Sharma Sanjib, Stello Dennis, Huber Daniel, Bland-Hawthorn Joss, \& Bedding Timothy R., 2017, ApJ, 835, 163S
\bibitem[\protect\citeauthoryear{Sch\"{o}nrich \& Binney}{2009a}]{Schonrich09} Sch\"{o}nrich R., Binney J., 2009a, MNRAS, 396, 203
\bibitem[\protect\citeauthoryear{Sch\"{o}lpokf et al.}{1998}]{Sch98} Sch\"{o}lpokf B., Smola A. J., M\"{u}ller K. R., 1998, Neural Comput., 10, 1299
\bibitem[\protect\citeauthoryear{Stello et al.}{2009}]{b8} Stello D. et al., 2009, ApJ, 700, 1589
\bibitem[\protect\citeauthoryear{Stello et al.}{2013}]{b9} Stello D. et al., 2013, ApJ, 765, L41
\bibitem[\protect\citeauthoryear{Stello et al.}{2015}]{b10} Stello D. et al., 2015, ApJ, 809, L3
\bibitem[\protect\citeauthoryear{Stello et al.}{2017}]{stello17} Stello D. et al., 2017, ApJ, 835, 83S
%\bibitem[\protect\citeauthoryear{Steinmetz et al.}{2006}]{Steinmetz06} Steinmetz M. et al., 2006, AJ, 132, 1645
\bibitem[\protect\citeauthoryear{Soderblom}{2010}]{Soderblom10} Soderblom D. R., 2010, ARA\&A, 48, 581
\bibitem[\protect\citeauthoryear{Tassoul}{1980}]{Tassoul1980} Tassoul M., 1980, ApJS, 43, 469
\bibitem[\protect\citeauthoryear{Takeda}{2007}]{Takeda2007} Takeda G., Ford Eric B., Sills Alison, et al., 2007, ApJS, 168, 297E
\bibitem[\protect\citeauthoryear{Ulrich}{1986}]{Ulrich1986} Ulrich R. K., 1986, ApJ, 306, L37
\bibitem[\protect\citeauthoryear{Viani et al.}{2017}]{Viani17} Viani Lucas S., Basu Sarbani, Chaplin William J., et al. 2017, ApJ, 843, 11
\bibitem[\protect\citeauthoryear{Vrard et al.}{2016}]{Vrard16} Vrard M., Mosser B., Samadi R., 2016, A\&A, 588A, 87V
\bibitem[\protect\citeauthoryear{Wang et al.}{2016}]{Wang16} Wang L., Wang W., Wu Y., et al., 2016, AJ, 152, 6
\bibitem[\protect\citeauthoryear{White et al.}{2011}]{White11} White T. R., Bedding T. R., Stello D., et al. 2011, ApJ, 743, 161
\bibitem[\protect\citeauthoryear{Wu et al.}{2011}]{Wu11} Wu Y. et al., 2011, RAA, 11, 924
%\bibitem[\protect\citeauthoryear{Wu et al.}{2014}]{Wu14} Wu Y., Du B., Luo A., Zhao Y., Yuan H., 2014, in Heavens A., Starck J.-L., Krone-Martins A., eds, Proc. IAU Symp. 306, Statistical Challenges in 21st Century Cosmology. Cambridge Univ. Press, Cambridge, p. 340
\bibitem[\protect\citeauthoryear{Wu et al.}{2017}]{Wu17} Wu Y. Q., et al., 2017, RAA, 17, 5
\bibitem[\protect\citeauthoryear{Xiang et al.}{2015a}]{Xiang15a} Xiang M. S. et al., 2015, MNRAS, 448, 822
\bibitem[\protect\citeauthoryear{Xiang et al.}{2015b}]{Xiang15b} Xiang M. S. et al., 2015, RAA, 15, 1209X
\bibitem[\protect\citeauthoryear{Xiang et al.}{2017a}]{Xiang17a} Xiang M. S. et al., 2017, MNRAS, 464, 3657
\bibitem[\protect\citeauthoryear{Xiang et al.}{2017b}]{Xiang17b} Xiang M. S. et al., 2017, ApJS, 232, 2
\bibitem[\protect\citeauthoryear{Xiang et al.}{2017c}]{Xiang17c} Xiang M. S. et al., 2017, MNRAS, 467, 1890X
\bibitem[\protect\citeauthoryear{Yanny et al.}{2009}]{Yanny09} Yanny B., Newberg H. J., Johnson J. A., et al., 2009, ApJ, 700, 1282
\bibitem[\protect\citeauthoryear{Yu et al.}{2016}]{Yu16} Yu Jie., Huber D., Bedding T. R., et al., 2016, MNRAS, 463, 1297
\bibitem[\protect\citeauthoryear{Yu et al.}{2017}]{Jie} Yu Jie., et al., 2017, submitted to the ApJ.
\bibitem[\protect\citeauthoryear{Yuan et al.}{2015}]{Yuan15} Yuan H.-B. et al., 2015, MNRAS, 448, 855
\bibitem[\protect\citeauthoryear{Zhao et al.}{2012}]{Zhao12} Zhao G., Zhao Y. H., Chu Y. Q., et al., 2012, RAA, 12, 723

\end{thebibliography}
\end{document}